# Data-driven design of fault diagnosis for three-phase PWM rectifier using random forests technique with transient synthetic features

Lei Kou, Chuang Liu, Guo-wei Cai, Jia-ning Zhou, Quan-de Yuan

## Abstract

A three-phase pulse-width modulation (PWM) rectifier can usually maintain operation when open-circuit faults occur in insulated-gate bipolar transistors (IGBTs), which will lead the system to be unstable and unsafe. Aiming at this problem, based on random forests with transient synthetic features, a data-driven online fault diagnosis method is proposed to locate the open-circuit faults of IGBTs timely and effectively in this study. Firstly, by analysing the open-circuit fault features of IGBTs in the three-phase PWM rectifier, it is found that the occurrence of the fault features is related to the fault location and time, and the fault features do not always appear immediately with the occurrence of the fault. Secondly, different data-driven fault diagnosis methods are compared and evaluated, the performance of random forests algorithm is better than that of support vector machine or artificial neural networks. Meanwhile, the accuracy of fault diagnosis classifier trained by transient synthetic features is higher than that trained by original features. Also, the random forests fault diagnosis classifier trained by multiplicative features is the best with fault diagnosis accuracy can reach 98.32%. Finally, the online fault diagnosis experiments are carried out and the results demonstrate the effectiveness of the proposed method, which can accurately locate the open-circuit faults in IGBTs while ensuring system safety.

## 1 Introduction

Nowadays, the power system has faced serious harmonic pollution problems due to the increasing usage of different electronic equipment [1]. A three-phase pulse-width modulation (PWM) rectifier has been attractive equipment to deal with such a problem because of its excellent performance and potential advantages such as a well-regulated DC-link voltage, nearly sinusoidal input currents, and tunable high-input power factor [2]. The three-phase PWM rectifiers have been widely used in interfacing distributed generation systems, AC–DC grid, drives of electrical motors, and other fields [3-5]. Therefore, it is very important to study the fault diagnosis of three-phase PWM rectifiers to improve the stability of the whole power system.

Although there are various measures to improve the stability of three-phase PWM rectifier equipment, a fault is still inevitable [6]. Open-circuit faults and short-circuit faults are the most common faults of the three-phase PWM rectifier. Since short-circuit faults in insulated-gate bipolar transistors (IGBTs) are destructive and easy to cause over-current, but the protection of short-circuit fault is also completed by a standard protection circuit [7, 8]. A method of turning short-circuit fault into an open-circuit fault by fast fuse was put forward in [9], thereby reducing the impact of faults. Open-circuit faults in IGBTs are usually caused by over-current burnout or caused by open-circuits of driven signals etc. [10, 11]. The open-circuit faults in IGBTs are less serious than that of short-circuit faults, which will not cause the system to stop immediately [12, 13]. Nevertheless, long-term fault operation is likely to lead secondary faults to the equipment system, which will result in higher maintenance costs [14]. The circuit parameters of three-phase PWM rectifiers, such as the equivalent resistance and inductance, vary with temperature and other environmental conditions. Therefore, it is extremely hard to establish fault mathematical models of power electronic devices [15].

Since the fault mathematical model is difficult to apply in practical application, meanwhile the fault diagnosis methods based on machine learning algorithms rely less on mathematical models, only historical experimental data or simulation data being used. Therefore, the methods based on machine learning algorithms are concerned by more and more scholars [16, 17]. A diagnosis method was proposed in [18] to detect and locate the open-circuit faults in IGBTs, the three-phase output currents were preprocessed by the discrete wavelet transform, which found that the values of Euclidean distance under open-circuit faults will be smaller than that under normal conditions, and then faults can be detected. Based on modifying established Park's vector method and normalised DC method, a fault detection method was proposed for open-circuit faults in IGBTs of the three-phase AC/DC PWM converter in [19]. A cost function and wavelet transform-based method was proposed in [20] for the interturn fault diagnosis of the permanent magnet synchronous machine. An artificial neural network (ANN)-based fault protection method was proposed in [21] for the modular multilevel converter, in which the input signals are preprocessed by discrete wavelet transform. In recent years, most of the related research for open-circuit fault diagnosis in three-phase DC/AC inverter has been developed, but the research on fault diagnosis of the three-phase PWM rectifier is relatively few, especially the three-phase four-wire PWM rectifier (hereinafter referred to as the three-phase PWM rectifier) [22]. Fault diagnosis in rectifier has its unique features, in comparison with open-circuit fault diagnosis in the inverter. The equipped diode will continually work as a rectifier component when open-circuit faults occur in one or more IGBTs. Except that the performance degrades such as output voltage fluctuation and current harmonics, the system will not crash immediately [23, 24]. Therefore, it is necessary to study the early fault features of three-phase PWM rectifiers to get rid of safety hazards in time.

Random forests (RFs) algorithm is applicable in regression and classification, which has excellent behaviour in the fields such as pattern recognition, condition monitoring, system control, and so forth [25-29]. Based on extracting features from the differential measured current data, a RFs-based fault discrimination method was proposed for power transformer in [27], which can effectively distinguish between internal faults and disturbances. Based on XGBoost and RFs, a data-driven wind turbine fault detection model was presented in [28], which can protect against over-fitting when dealing with multidimensional data.

Several different fault diagnosis methods for power electronic devices are compared in Table 1. Research Studies have achieved good performance in their respective fields, which gave us great inspiration and experience. Compared with the model-based fault diagnosis method, the data-driven method needs a longer time, but it has less dependence on the model. Also, unlike the methods in [10, 12, 13], it does not need to set complex thresholds. Most fault diagnosis methods for power electronic devices have different research backgrounds, some of them need expensive equipment, complex parameters, knowledge, or fault models. Generally, the magnitude of fault-related feature in the transient state is greater than the steady-state fluctuation [31]. Meanwhile, three-phase four-wire PWM rectifiers have the function of unbalanced-power compensation and active-power filter, which play an important role in the field of low-voltage AC–DC grid. Therefore, a new online fault diagnosis data-driven method based on RFs is proposed in this paper, and takes the three-phase four-wire PWM rectifier as the example. Of course, the proposed method is also applicable to most of the power-electronics conversion devices. The open-circuit faults in IGBTs of the three-phase PWM rectifier are deeply studied, and the synthetic features method is used to improve the RFs fault diagnosis algorithm. The fault diagnosis classifier is trained by the transient synthetic features, which has a strong generalisation ability for the open-circuit faults in IGBTs. Furthermore, a low-cost multi-time scale fault diagnosis method is proposed, which can locate and detect the fault location on the premise of ensuring the system safe. Generally, there are few features that can be collected in power electronic devices, but few studies consider to expand the features. Therefore, the main contributions of this paper are summarised as follows:

(i) A method of synthetic features is proposed to improve the accuracy of a data-driven algorithm, where the fault classifier trained by multiplicative features is better than that of additive features.

**Table 1.** Comparison of different methods

| Methods | Circuit structure | Comments |
| --- | --- | --- |
| robust observer-based method [10] | three-phase voltage-source inverters | implemented with DSPACE DS1103, very expensive |
| time-domain analysis of state observer residual [12] | four-phase interleaved buck converter | rely on models, complex parameters |
| state observers and knowledge of fault behaviours based [13] | modular multilevel converter | only diagnosis for single IGBT, requirement for knowledge of topology |
| current similarity based [18] | two-level three-phase voltage-source inverters | complex parameters |
| modify Park's vector method and normalised DC current method [19] | three-phase AC/DC PWM converter | requirement for knowledge of topology |
| modify the topology [30] | neutral-point-clamped (NPC) five-level inverter | requirement for knowledge and models |
| proposed method | three-phase PWM rectifier | independent of fault model |

(ii) RFs are trained by the transient synthetic features, which are easier to get a more accurate fault diagnosis classifier for power electronics circuits.
(iii) A low-cost multi-time scale fault diagnosis method is proposed, which can accurately realise the fault location for power electronics circuits.

Fig. 1 shows the diagram of the online fault diagnosis system for the three-phase PWM rectifier, and the main work that has been done is described as following: Section 2 analyses the fault features of open-circuit faults in IGBTs of the three-phase PWM rectifier, and the actual experiments are adopted to simulate the open-circuit faults and collect the fault samples; in Section 3, different fault diagnosis methods are compared and evaluated. In Section 4, the online fault diagnosis experiments are carried out to verify that the classifier can accurately locate the open-circuit faults' locations of IGBTs by transient synthetic features. The conclusion is presented in Section 5.

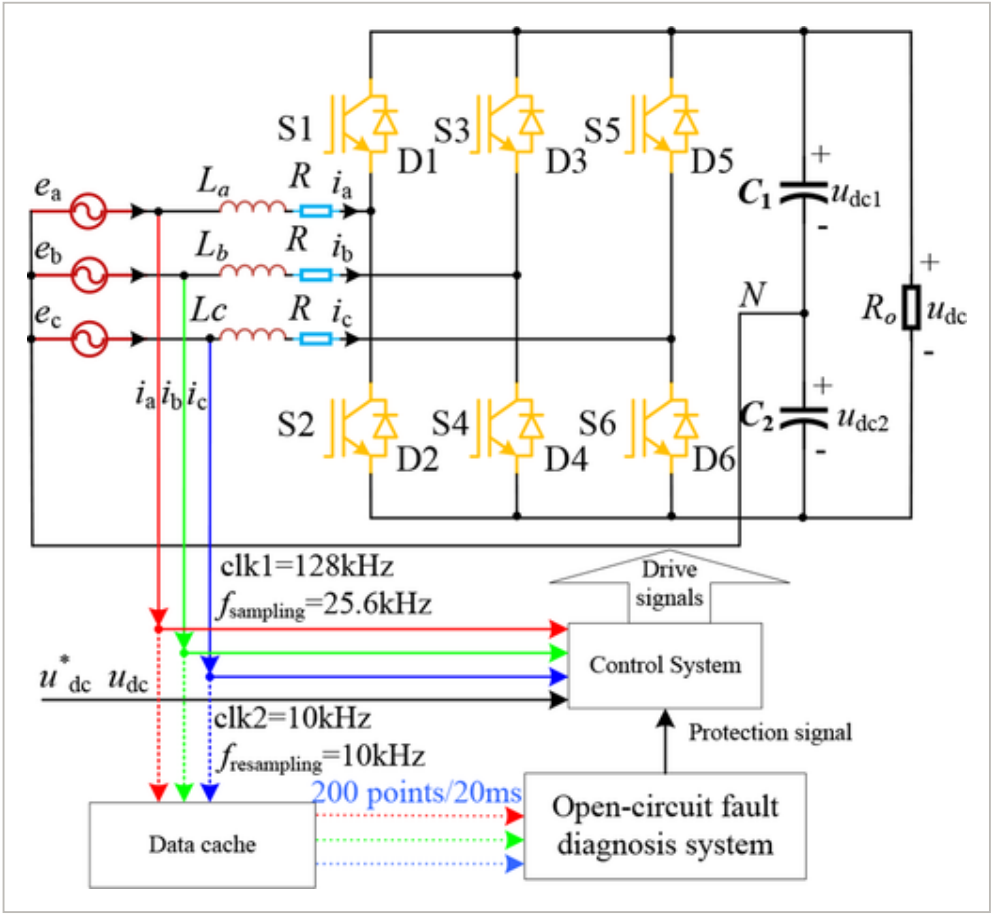

**Fig. 1**

Open in figure viewer | PowerPoint

*Diagram of online fault diagnosis system for three-phase PWM rectifier*

## 2 Fault data acquisition and features analysis

### 2.1 Fault data acquisition

To maintain consistency with the actual operation, the actual fault experiments are adopted to collect historical data under different fault conditions in this paper. The open-circuit faults in IGBTs are usually caused by the driven faults or the destruction of devices. Therefore, the open-circuit faults in IGBTs can be simulated by shutting down the control signals of IGBTs. Moreover, the main circuit topology of the three-phase PWM rectifier is shown in Fig. 2, where Fig. 2 shows the open-circuit faults in S1 and S3. For the proportional–resonant controller used in this paper, refer to [32], and the control block diagram of the three-phase PWM rectifier is shown in Fig. 3.

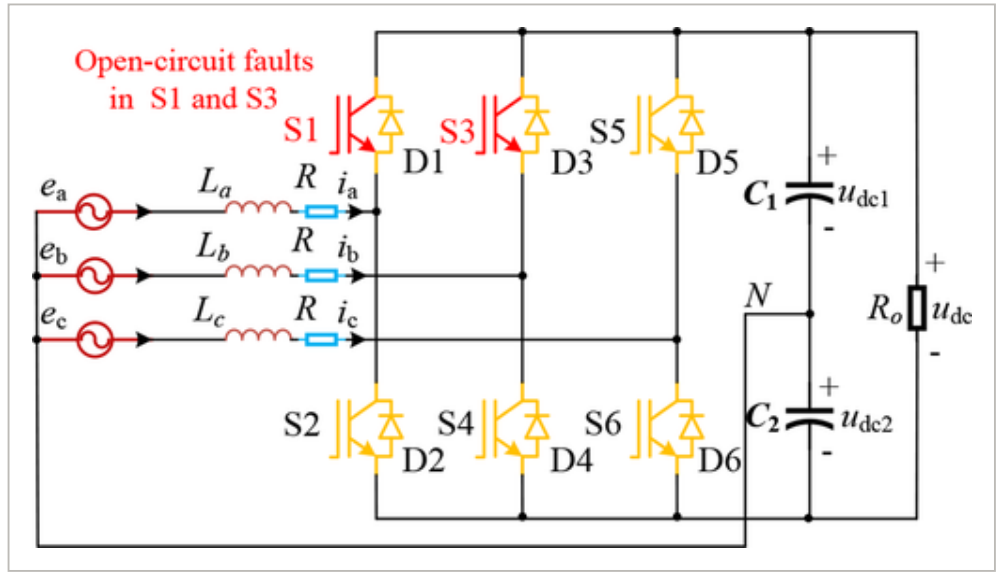

**Fig. 2**

Open in figure viewer | PowerPoint

*Three-phase PWM rectifier with open-circuit faults in S1 and S3*

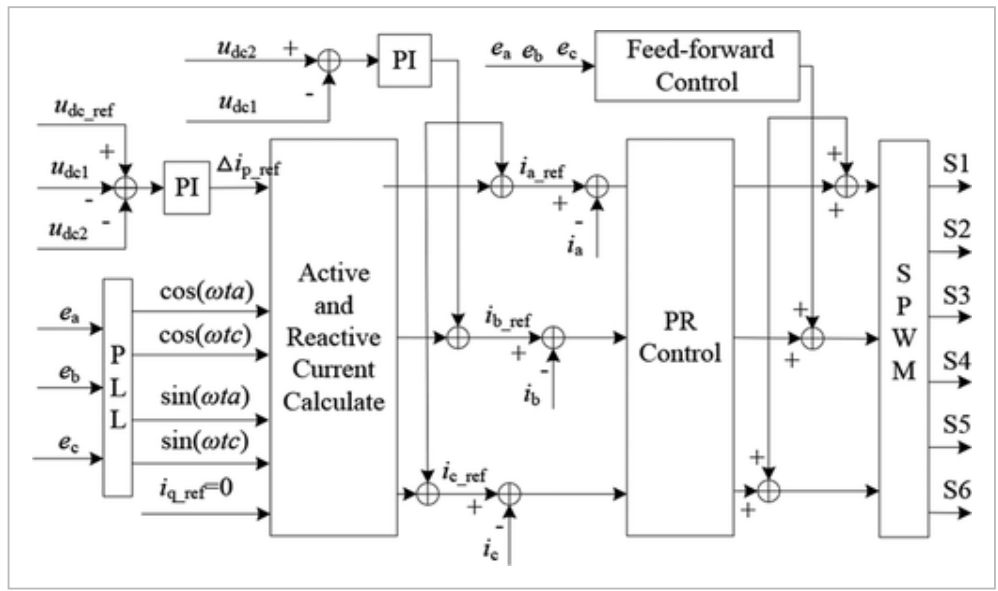

**Fig. 3**

Open in figure viewer | PowerPoint

*Control block diagram of three-phase PWM rectifier*

The circuit parameters of the experimental system are listed in Table 2, and the experimental device is shown in Fig. 4. Fig. 5 shows the waveform diagram of the phase current when open-circuit faults occur in IGBTs, and they are the experimental waveform of open-circuit faults in S1, S2, (S1 and S2), and (S1 and S3), respectively.

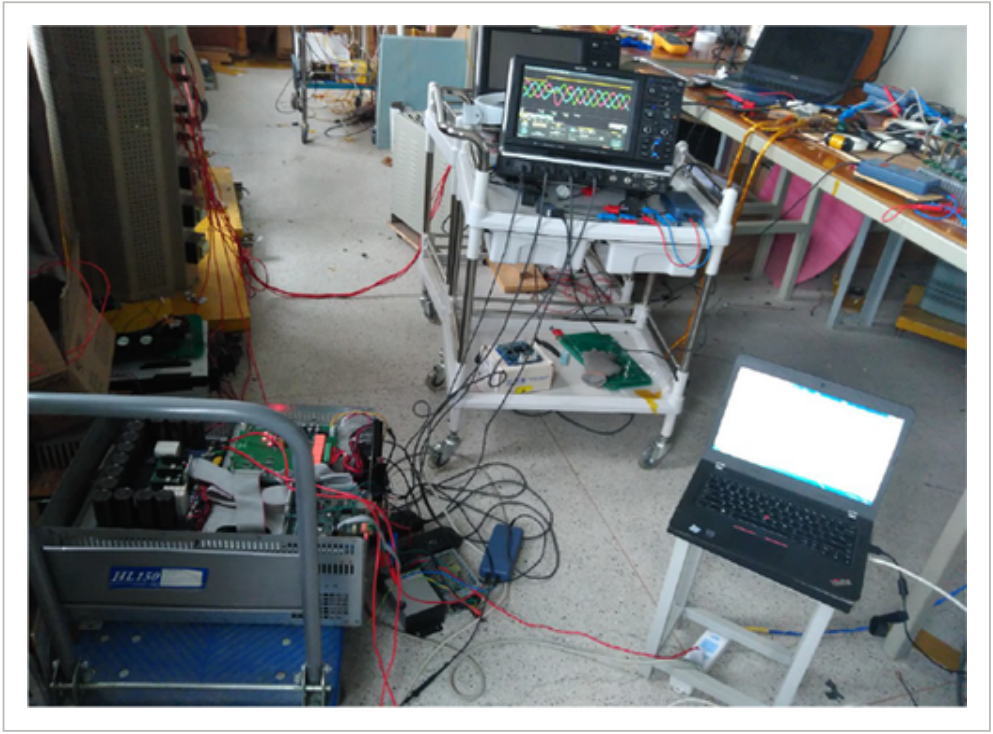

**Fig. 4**

Open in figure viewer | PowerPoint

*Experimental platform system*

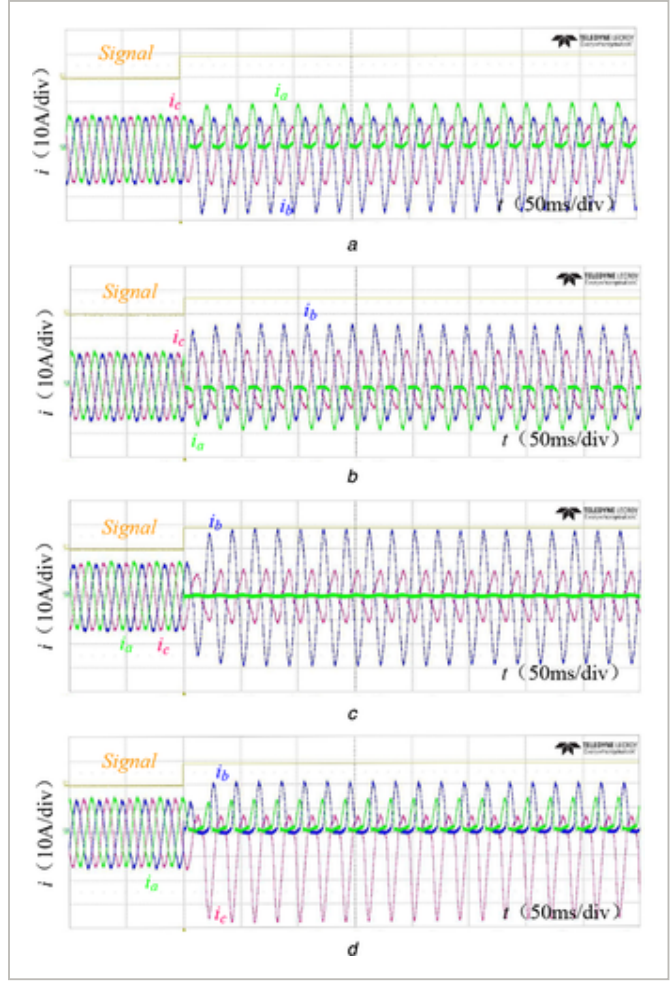

**Fig. 5**

Open in figure viewer | PowerPoint

*Experimental waveform*

*(a)* S1 open-circuit fault, *(b)* S2 open-circuit fault, *(c)* S1 and S2 open-circuit faults, *(d)* S1 and S3 open-circuit faults

**Table 2.** Circuit parameters of the experimental system

| Parameters | Value |
| --- | --- |
| input voltage | 40 V |
| output voltage | 100 V |
| filter inductance | 500 µH |
| DC-link capacitances | 7000 µF |
| voltage frequency | 50 Hz |
| IGBT | FS1501212ET4 |
| sampling frequency | 25.6 kHz |
| switching frequency | 12.8 kHz |

| Parameters | Value |
|---|---|
| load | 16 Ω |

## 2.2 Feature analysis of open-circuit faults in IGBTs

Fig. 6 shows the phase current waveform when the open-circuit fault occurs in a single insulated-gate bipolar transistor (IGBT) of a three-phase PWM rectifier. According to Fig. 6, when the open-circuit fault occurs in only one IGBT, the corresponding phase current changes sharply instantaneously, and the other two-phase currents will also appear in a certain distortion. The open-circuit faults in upper IGBTs will affect the negative half-wave of the corresponding phase current, and the open circuit faults in lower IGBTs will affect the positive half-wave.

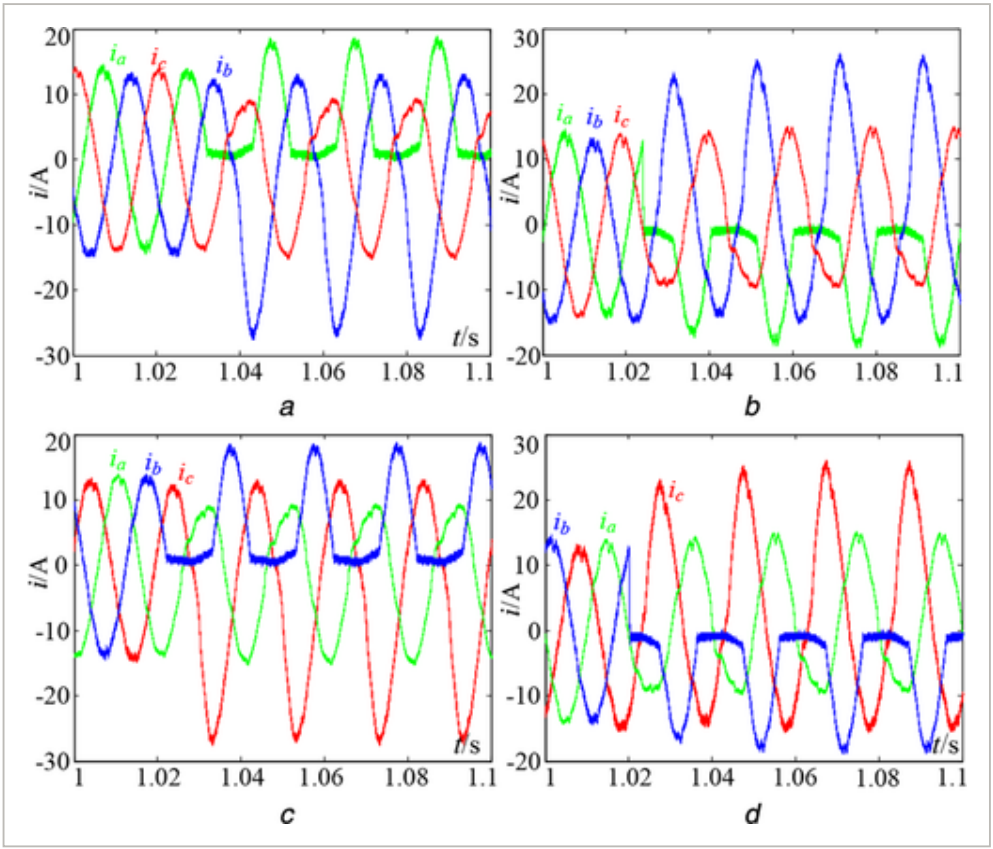

**Fig. 6**

Open in figure viewer | PowerPoint

*Waveform of open-circuit fault in single IGBT*

*(a)* S1 fault, *(b)* S2 fault, *(c)* S3 fault, *(d)* S4 fault

As shown in Fig. 7, taking the current $i_a$ as an example, the reason for the above phenomenon is that the current $i_a$ in the PWM rectifier consists of the following conducting routes.

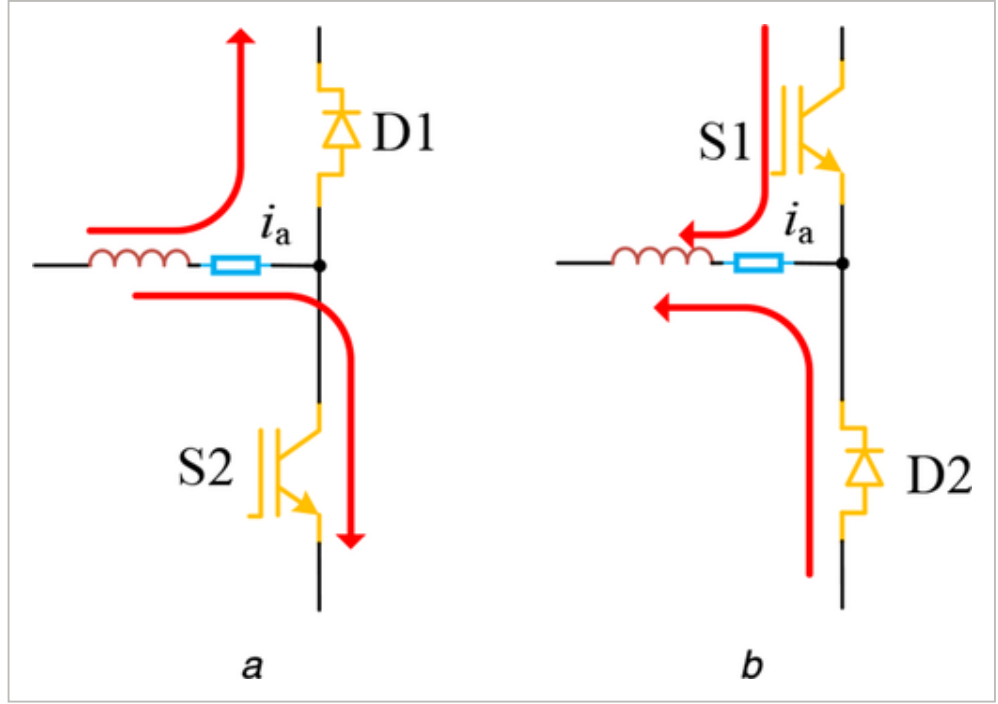

**Fig. 7**

*Conducting route under different directions of $i_a$*

*(a) $i_a \geq 0$, (b) $i_a < 0$*

When $i_a \geq 0$, S1 is turn-on and S2 is turn-off, $i_a$ flows through equipped diode D1, and then continued flow can be achieved. Also, when $i_a \geq 0$, S1 is turn-off and S2 is turn-on, $i_a$ flows through IGBT S2. In both cases, S1 is not involved in the current flow. Therefore, it has no effect on the PWM rectifier when the open-circuit fault in S1 and $i_a \geq 0$, but S2 will affect the current flow.

When $i_a < 0$, S1 is turn-on and S2 is turn-off, $i_a$ flows through IGBT S1, and then continued flow can be achieved. Also, when $i_a < 0$, S1 is turn-off and S2 is turn-on, $i_a$ flows through equipped diode D2. In both cases, S2 is not involved in the current flow. Therefore, it has no effect on the PWM rectifier when the open-circuit fault is in S2 and $i_a < 0$, but S1 will affect the current flow.

As shown in Fig. 8, Fig. 8a shows the fault waveform with open-circuit fault occurring simultaneously in IGBTs S1 and S2, and Fig. 8b shows the fault waveform when open-circuit fault occurs simultaneously in IGBTs S1 and S3. According to Figs. 6 and 8, it can be seen that the fault features do not always appear on the phase currents immediately when the fault occurs. S1 and S2 are IGBTs of the same bridge arm, when open-circuit faults occurred in S1 and S2 at the same time, their fault features will not show up at the same time, but each occupies half a cycle. S1 and S3 are distributed on the different arm, respectively, and the performance of the fault features is related to the direction of the three-phase current. As shown in Fig. 8b, since S1 and S3 are distributed on the different arm, respectively, the fault features are related to the direction of the three-phase phase current, and the order of features is S3 fault → S0 (normal state) → S1 fault → (S1 and S3 fault) → S3 fault.

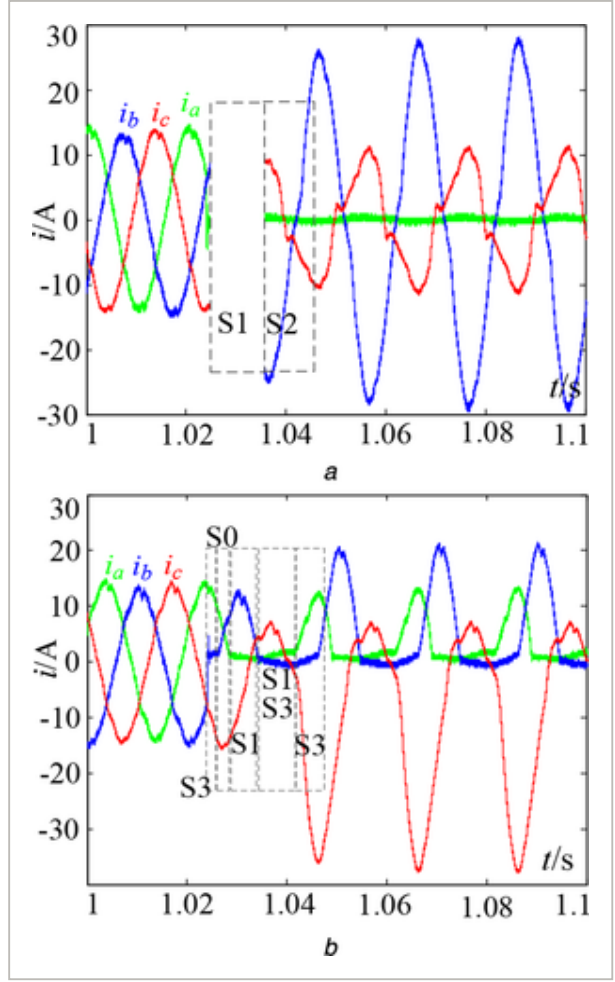

**Fig. 8**

Open in figure viewer | PowerPoint

*Waveform of open-circuit fault in IGBTs at the same time*

*(a)* S1 and S2 open-circuit faults, *(b)* S1 and S3 open-circuit faults

Fig. 9 shows the phase current waveform when open-circuit faults occur in multiple IGBTs. It can be seen that the fluctuation of phase current is more serious when open-circuit faults occur in multiple IGBTs. The three-phase PWM rectifier will enter an uncontrolled mode and the output voltage will drop severely only when open-circuit faults occur in all IGBTs. Therefore, the PWM rectifier may operate with faults when open-circuit faults occur in part IGBTs, which is easy to bring secondary faults to other equipment.

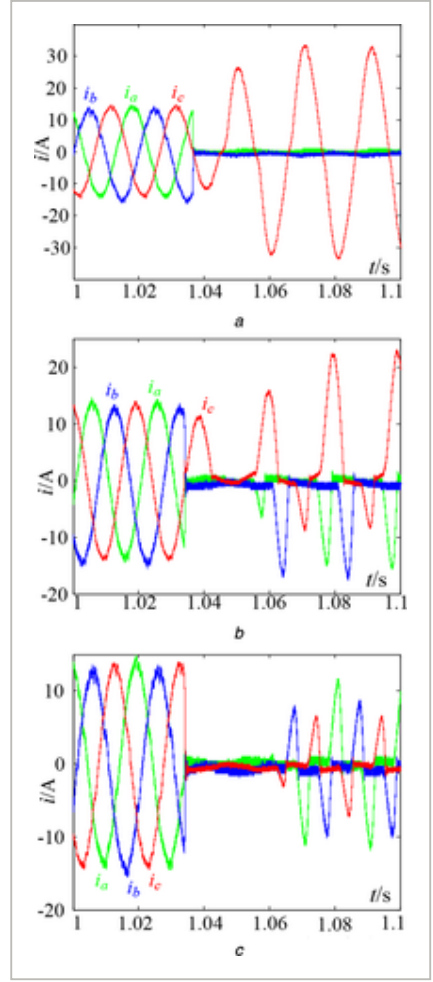

**Fig. 9**

*Waveform of open-circuit faults in multiple IGBTs*

*(a)* S1, S2, S3 and S4 faults, *(b)* S1, S2, S3, S4 and S5 faults, *(c)* S1, S2, S3, S4, S5 and S6 faults

According to Figs. 6-9, whether the three-phase PWM rectifier works in controlled or uncontrolled modes depends on the number and the locations of IGBT faults. In the case of IGBT faults, the current of the AC side in the PWM rectifier is not always zero but is a great extent to approach the sine wave, meanwhile, it is not the same as the ideal mathematical model.

## 3 Training and evaluation of fault diagnosis classifiers

In this section, different data-driven fault diagnosis methods are compared and evaluated, and the accuracies of fault diagnosis methods are improved by using synthetic features.

### 3.1 Training of fault diagnosis classifiers

The RFs algorithm, an ensemble classifier composed of many decision trees, was proposed by Breiman *et al.* [33], which can acquire strong predictive models, train rapidly and eliminate the issue of overfitting. In addition, support vector machine (SVM) and ANN algorithm are often used for fault diagnosis [21, 34]. Therefore, this paper makes a comparative analysis of RFs, ANN, and SVM algorithm, and for the training process of ANN and SVM algorithm refer to [21, 35]. To facilitate the presentation and convenience for the training of data-driven fault diagnosis classifiers, the faults' locations are encoded as shown in Table 3, when the open-circuit faults in the IGBT $S_k$ ($k$ =1–6), the output codes $d_k$ ($k$ =1–6) = 1.

**Table 3.** Fault IGBT and output codes

| Fault IGBT | Output codes | | | | | |
|---|---|---|---|---|---|---|
| | d1 | d2 | d3 | d4 | d5 | d6 |
| normal state | 0 | 0 | 0 | 0 | 0 | 0 |
| S1 | 1 | 0 | 0 | 0 | 0 | 0 |
| S2 | 0 | 1 | 0 | 0 | 0 | 0 |
| S3 | 0 | 0 | 1 | 0 | 0 | 0 |
| S4 | 0 | 0 | 0 | 1 | 0 | 0 |
| S5 | 0 | 0 | 0 | 0 | 1 | 0 |
| S6 | 0 | 0 | 0 | 0 | 0 | 1 |
| S1 and S2 | 1 | 1 | 0 | 0 | 0 | 0 |
| S1 and S3 | 1 | 0 | 1 | 0 | 0 | 0 |

In the training progress of data-driven fault diagnosis classifiers, given $(x_i, y_i)$ pairs $i = 1, 2, \ldots, N$, where $x_i$ is the input feature $(i_a, i_b, i_c)$, $y_i$ is the response for classification, i.e. the output label.

To reduce the influence from the numerical value of fault samples on the training accuracy of fault diagnosis classifier, the fault samples are normalised to $[x'_{min}, x'_{max}]$, where $x'_{min} = -1$, $x'_{max} = 1$. The normalised expression is

$$x' = \begin{cases} \frac{(x'_{max} - x'_{min})(x - x_{min})}{(x_{max} - x_{min})} + x'_{min}, & x_{max} \neq x_{min} \\ x'_{min}, & x_{max} = x_{min} \end{cases} \quad (1)$$

where $(i_a, i_b, i_c)$ are the fault samples, $x_{min}$ and $x_{max}$ are the maximum and minimum values of the fault samples in the same group, respectively, and $x'$ is the normalised data.

Assumed that a training set is given by $X = x_{1,2,\ldots,n}$ with response $Y = y_{1,2,\ldots,n}$, bagging will select a random sample to replace the training set and repeat $N$ times, and then fit trees to these samples, and $h(x)$ will be trained every time. RFs consist of multiple $h(x)$, where $h(x)$ is a classifier with a tree structure. After training, the mature fault diagnosis classifier will be established by taking the majority vote from $N$ decision trees. For the steps of the general process for RFs generation refer to [36], for the voting rules refer to [37], and for the description of Gini index refer to [38]. Fig. 10 shows the RFs algorithm with synthetic features, and Fig. 11 shows the flow chart of RFs with synthetic features. In the training process of a fault diagnosis classifier, it takes a certain amount of time and calculation. Also, in the process of fault diagnosis, it almost does not consume time and computational effort to locate the fault location with the help of mature fault diagnosis classifiers.

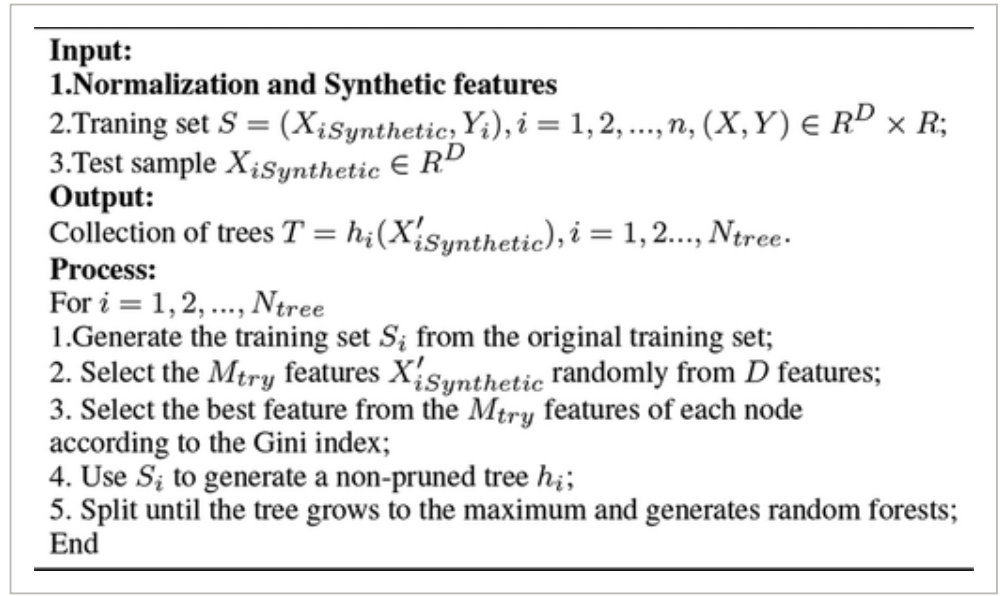

**Fig. 10**

RFs algorithm with synthetic features

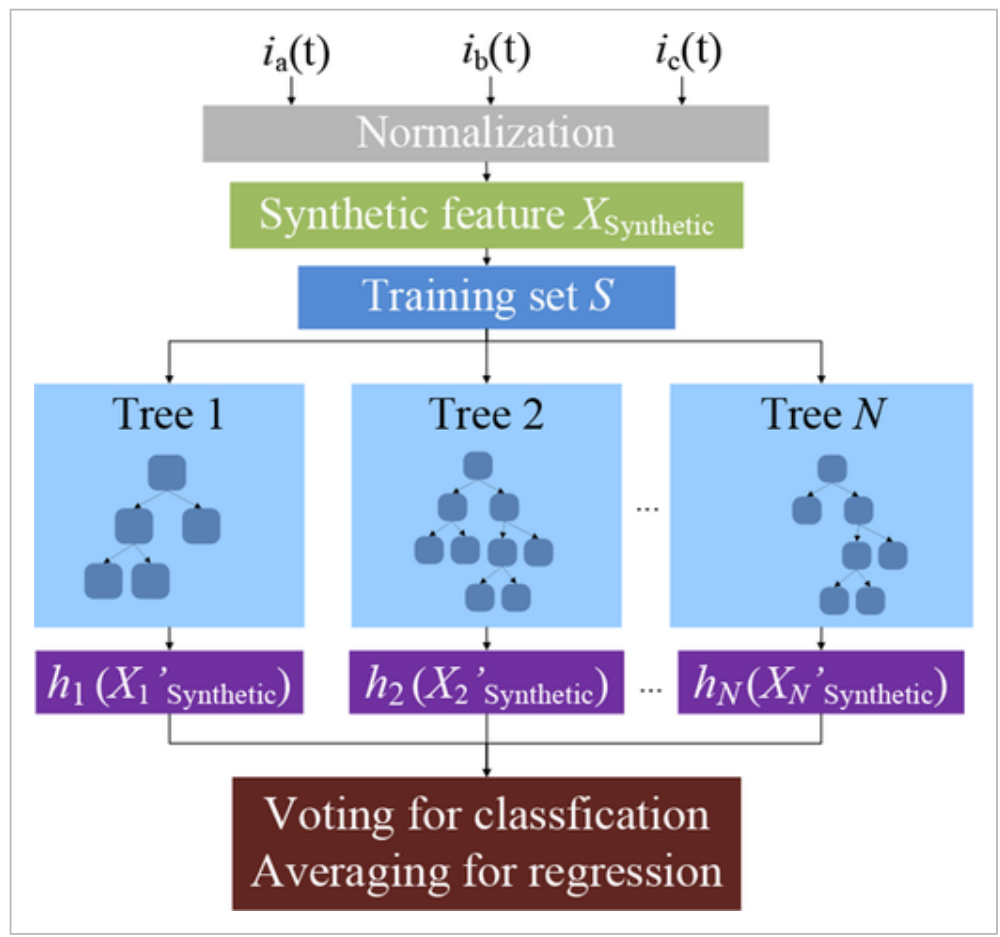

**Fig. 11**

*Flow chart of RFs with synthetic features*

Assuming that the relationship between input $X$ and output $Y$ is $Y = f(X)$, there are $n$ different weak classifiers $h_1(X), h_2(X), \ldots, h_N(X)$, the expected error of each classifier can be defined as

$$\begin{aligned} \Re(h_n) &= E_X\left[(h_n(X) - f(X))^2\right] \\ &= E_X\left[\varepsilon_n^2(X)\right] \end{aligned} \tag{2}$$

where $\varepsilon_n(X) = h_n(X) - f(X)$ is the nth error of the input sample $X$. Then the average error of all classifiers can be expressed as

$$\begin{aligned} \bar{\Re}(h_n) &= \frac{1}{N} \sum_{n=1}^{N} E_X\left[(h_n(X) - f(X))^2\right] \\ &= \frac{1}{N} \sum_{n=1}^{N} E_X\left[\varepsilon_n^2(X)\right] \end{aligned} \tag{3}$$

The commonly used ensemble method of ensemble learning mainly includes the direct average method and weighted average method, where the direct average method is also known as the voting method, and the ensemble classifier can be defined as

$$H(x) = \frac{1}{N} \sum_{n=1}^{N} h_n(X) \tag{4}$$

Then the expected error of the ensemble classifier can be expressed as

$$\begin{aligned} \Re(H) &= E_X\left[\left(\frac{1}{N} \sum_{n=1}^{N} h_n(X) - f(X)\right)^2\right] \\ &= \frac{1}{N^2} E_X\left[\left(\sum_{n=1}^{N} \varepsilon_n(X)\right)^2\right] \\ &= \frac{1}{N^2} E_X\left[\sum_{n=1}^{N}\sum_{m=1}^{N} \varepsilon_n(X)\varepsilon_m(X)\right] \\ &= \frac{1}{N^2} \sum_{n=1}^{N}\sum_{m=1}^{N} E_X\left[\varepsilon_n(X)\varepsilon_m(X)\right] \end{aligned} \tag{5}$$

where $E_X[\varepsilon_n(X)\varepsilon_m(X)]$ is the correlation between the errors of two different weak classifiers. If the errors of all different weak classifiers are not related, then $E_X[\varepsilon_n(X)\varepsilon_m(X)] = 0$ for any $m \neq n$. If all the errors are the same, then $\varepsilon_n(X) = \varepsilon_m(X)$ for any $m \neq n$. Since $\varepsilon_n(X) \geq 0$ for any weak classifiers, therefore, the following expression should be satisfied:

$$\bar{\Re}(h) \geq \Re(H) \geq \frac{1}{N}\bar{\Re}(h) \tag{6}$$

Therefore, the expected error of the ensemble classifier is between $(1/N)\bar{\Re}(h)$ and $\bar{\Re}(h)$. According to the above deduction, the greater difference between the weak classifiers, the better the ensemble classifier. Therefore, compared with SVM and ANN, RFs algorithm can obtain higher classification accuracy more easily.

Generally, the more features, the more independent trees that can be trained. When the input features are fewer, it is easy to lead to fewer types of trees in the RFs, and the classification performance of the fault diagnosis classifier will also be affected. Therefore, synthetic features are an effective strategy to extend new features for learning better models. The way of synthesis features mainly includes a feature multiplied with one or more other features (including itself), or a feature divided by other features, or the continuous feature divided into several intervals. The original features are. ., the additive features are
$(i_a, i_b, i_c, i_a + i_b, i_a + i_c, i_a + i_b + i_c, i_a - i_b, i_a - i_c, i_a - i_b - i_c)$, and the multiplicative features are
$(i_a, i_b, i_c, i_a * i_b, i_a * i_c, i_b * i_c, i_a * i_b * i_c, i_a / i_b, i_a / i_c, i_b / i_c)$. In this paper, the original features, additive features, and multiplicative features are used to train the RFs, ANN, and SVM algorithms, respectively.

The parameter for RFs is determined by the cross-validation method, and the number of trees in RFs is usually determined by experimental experience. As shown in Fig. 12, the accuracies of their classification vary with the number of trees. It is obvious that the classifier trained by multiplicative features has higher accuracy than that trained by additive features and original features, where the original features exhibited the lowest accuracy. Also, their best numbers of trees, accuracy, and training time are shown in Table 4, where 16,800 fault samples of each fault condition were used for training, 7200 fault samples of each fault condition for testing, and all the fault samples are selected from actual fault experiments.

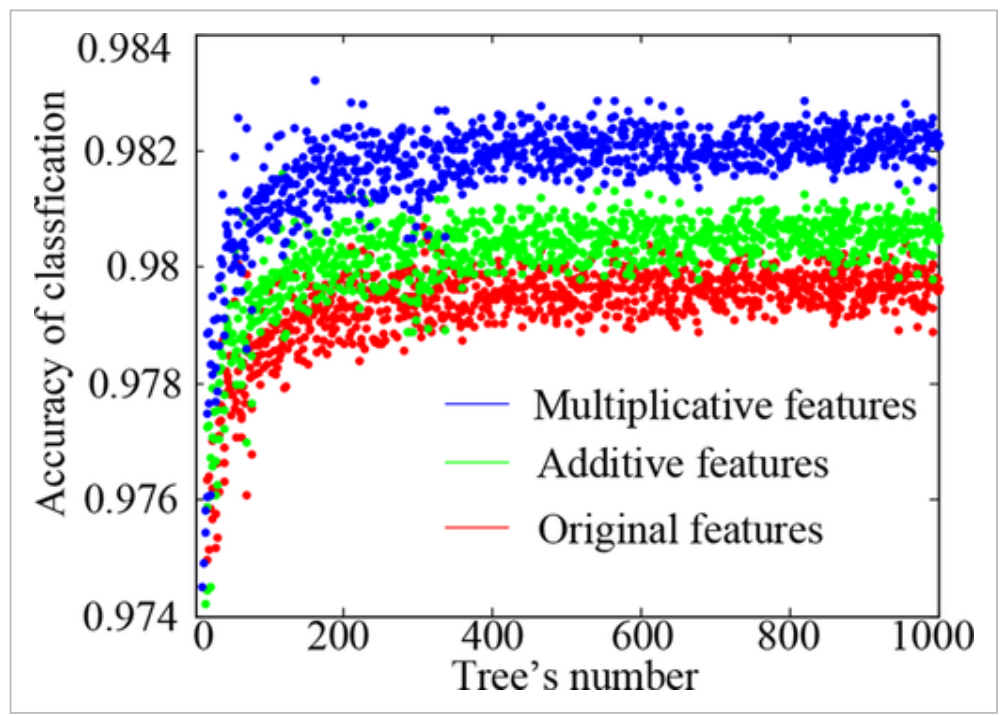

**Fig. 12**

Open in figure viewer | PowerPoint

*Influence of decision tree and features on performance in RFs*

**Table 4.** Comparison of performance

| Methods | Best tree's number | Classification accuracy | Training time |
|---|---|---|---|
| RFs + original features | 305 | 0.9807 | 4.70 min |
| RFs + additive features | 116 | 0.9816 | 3.97 min |
| RFs + multiplicative features | 159 | 0.9832 | 3.80 min |
| ANN + original features | — | 0.9631 | 50.26 min |
| ANN + additive features | — | 0.9643 | 48.86 min |
| ANN + multiplicative features | — | 0.9674 | 48.32 min |
| SVM + original features | — | 0.9622 | 53.62 min |
| SVM + additive features | — | 0.9635 | 52.72 min |
| SVM + multiplicative features | — | 0.9656 | 48.45 min |

According to Table 4, among the fault diagnosis methods, the training time of RFs with multiplicative features is the shortest, and the classification accuracy is the highest. In addition, the classification accuracy of RFs is higher than that of SVM, and the training time of RFs is shorter than that of SVM. Meanwhile, the performance of RFs algorithm with multiplicative features is the best.

## 3.2 Evaluation of fault diagnosis classifiers

Table 5 shows some samples and diagnosis output results. The mature diagnosis classifier outputs the diagnosis result when the transient three-phase currents are input, and then locate the faults of IGBTs.

**Table 5.** Some samples and diagnosis output results

| Fault IGBT | $i_a$ | $i_b$ | $i_c$ | Actual output | Target output |
|---|---|---|---|---|---|
| S0 | 0 | −14.280 | 13.615 | 000,000 | 000,000 |
| S1 | −0.332 | 7.638 | 2.324 | 100,000 | 100,000 |
| S2 | −2.324 | −13.6158 | 14.2800 | 010,000 | 010,000 |
| S3 | −12.287 | 0.664 | 13.947 | 001,000 | 001,000 |
| S1 and S3 | 0.664 | −0.332 | 5.645 | 101,000 | 101,000 |
| S1 and S4 | 0.012 | −3.976 | −3.869 | 100,100 | 100,100 |
| S1 and S5 | 4.660 | 0.002 | 1.012 | 100,010 | 100,010 |

Table 6 shows the comparison of different fault diagnosis classifiers' diagnosis accuracy. To calculate accuracies of ANN, SVM, and RFs, tests for different data rather than the training dataset are used. The datasets of other open-circuit faults tell a similar story, so only some results on two datasets are listed in Table 6.

**Table 6.** Diagnosis accuracy of different fault diagnosis classifiers

| Fault IGBT | RFs | | | SVM | | | ANN | |
|---|---|---|---|---|---|---|---|---|
| | Original features | Additive features | Multiplicative features | Original features | Additive features | Multiplicative features | Original features | Additive features |
| normal | 0.9769 | 0.9773 | 0.9802 | 0.9632 | 0.9654 | 0.9678 | 0.9682 | 0.9691 |
| S1 | 0.9673 | 0.9788 | 0.9846 | 0.9502 | 0.9512 | 0.9518 | 0.9512 | 0.9625 |
| S2 | 0.9842 | 0.9843 | 0.9848 | 0.9685 | 0.9702 | 0.9715 | 0.9672 | 0.9683 |
| S3 | 0.9832 | 0.9873 | 0.9882 | 0.9603 | 0.9613 | 0.9652 | 0.9660 | 0.9703 |
| S4 | 0.9986 | 0.9995 | 0.9997 | 0.9706 | 0.9734 | 0.9762 | 0.9812 | 0.9836 |
| S5 | 0.9908 | 0.9928 | 0.9935 | 0.9623 | 0.9657 | 0.9684 | 0.9744 | 0.9758 |
| S6 | 0.9965 | 0.9978 | 0.9986 | 0.9675 | 0.9716 | 0.9748 | 0.9783 | 0.9806 |
| S1 and S3 | 0.9636 | 0.9632 | 0.9675 | 0.9579 | 0.9623 | 0.9654 | 0.9551 | 0.9568 |
| S1 and S4 | 0.9705 | 0.9721 | 0.9721 | 0.9572 | 0.9583 | 0.9624 | 0.9608 | 0.9635 |
| S1 and S5 | 0.9642 | 0.9648 | 0.9648 | 0.9526 | 0.9578 | 0.9626 | 0.9528 | 0.9547 |
| S2 and | 0.9804 | 0.9812 | 0.9812 | 0.9619 | 0.9632 | 0.9656 | 0.9636 | 0.9656 |

According to Table 6, it can be found that the diagnosis accuracies of fault diagnosis classifiers trained by RFs are higher than those of ANN and SVM. It is obvious that the fault diagnosis classifier trained by synthetic features has higher accuracy than the classifier trained by original features. Compared to the classifiers trained by the other two features, the performance of the classifier trained by multiplicative features is the best, and the diagnosis accuracy is relatively balanced for most fault states. The RFs with multiplicative features offer extremely promising discrimination accuracy, which gives >98% average accuracy for different types of faults. Compared with the ANN and SVM algorithms, not only the actual computational quantity of the RFs algorithm is smaller, but also the diagnosis accuracy is higher.

# 4 Online fault diagnosis experiments

To evaluate the feasibility of the proposed method, the online fault diagnosis experiments for the three-phase PWM rectifier were performed. Fig. 1 shows the diagram of the online fault diagnosis system for the three-phase PWM rectifier. Since most artificial intelligence algorithms need to run on computers, a multi-time scale fault diagnosis method was adopted on the basis of the system safe. As shown in Fig. 13, the fault diagnosis system consists of the industrial computer and field-programmable gate array (FPGA) controller, where the bottom controller for open-circuit faults monitoring and protection is FPGA, and online fault diagnosis system is running on the industrial controlled computer. The industrial computer can be replaced by an ARM chip, the RFs can run on an ARM chip. As shown in Fig. 1, the sampling clock and sampling frequency of the closed-loop control system are 128 and 25.6 kHz, respectively. Also, the current signals are re-sampled with a clock of 10 kHz and a resampling frequency of 10 kHz. Therefore, it only sends 200 points every period (20 ms) to the industrial controlled computer to reduce data transmission pressure. The online fault diagnosis system can output 200 diagnosis results every period, and the 200 diagnosis results jointly decide the fault location. Once the open-circuit fault is detected, the online fault diagnosis system will send a protection signal to the FPGA controller to turn off all IGBT control signals.

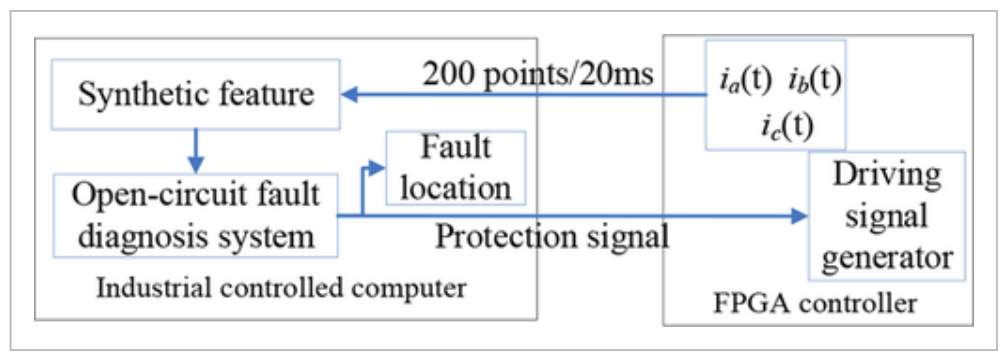

**Fig. 13**

*Diagram of data exchange between industrial computer and FPGA controller*

In this paper, S1 fault, S1 and S3 fault modes are taken as examples to illustrate the effectiveness of the proposed method. As shown in Fig. 14, once the open-circuit fault is detected, the FPGA controller will immediately turn off all IGBTs to protect the system. This method can locate the fault location while ensuring the system safe, and the diagnosis results are shown in Fig. 15.

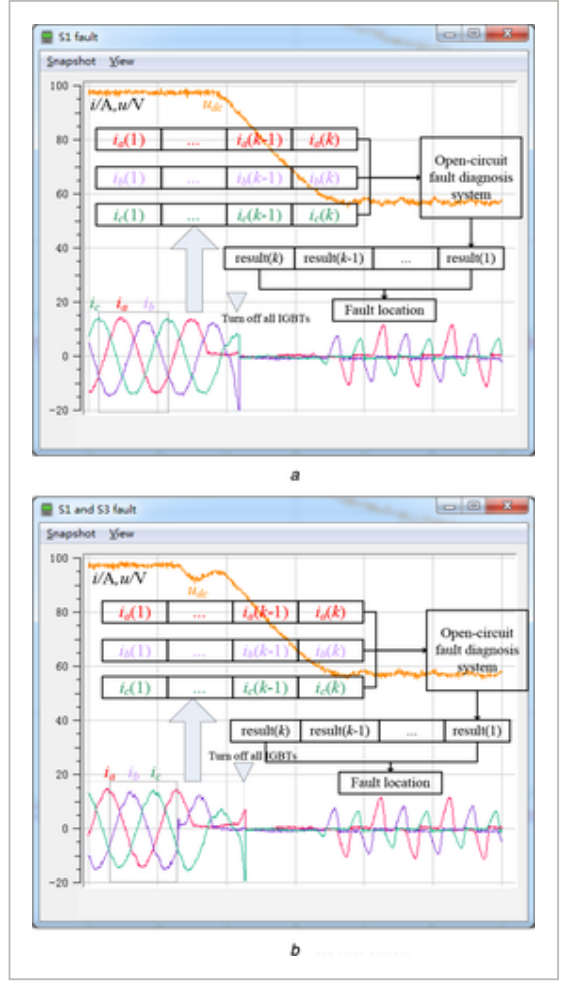

**Fig. 14**

Fault diagnosis experiments

*(a)* S1 open-circuit fault, *(b)* S1 and S3 open-circuit faults

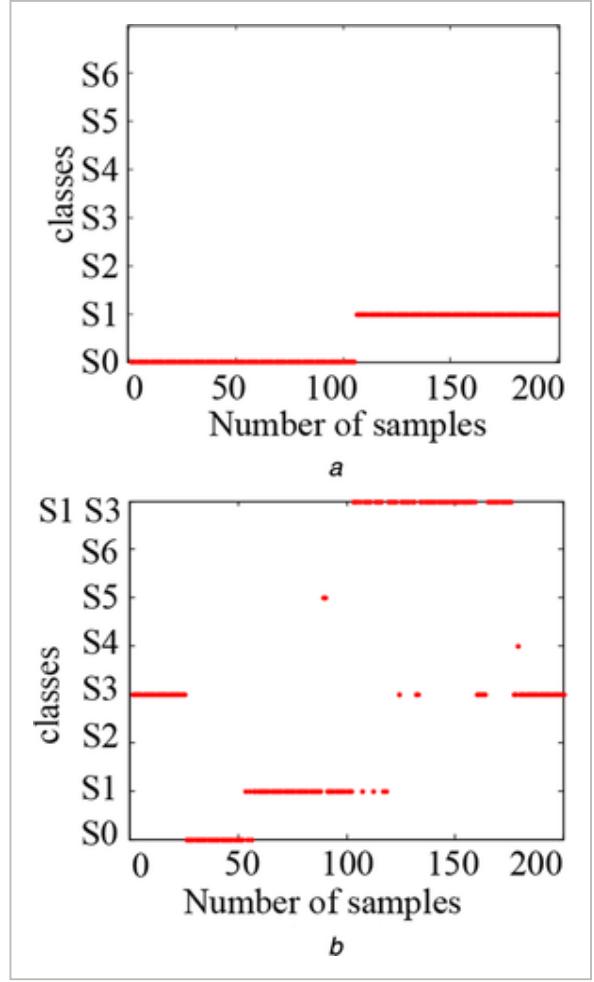

**Fig. 15**

Open in figure viewer | PowerPoint

*Fault diagnosis results*

*(a)* S1 fault diagnosis results, *(b)* S1 and S3 fault diagnosis results

Fig. 15 shows the diagnosis results of 200 samples. Fig. 15a shows the diagnosis result of an open-circuit fault in S1, in which the earlier parts of the diagnosis results are diagnosed as a normal state, for the reason that the fault features of S1 are not display immediately, the latter part is diagnosed as S1 fault, and finally the location of open-circuit fault is in IGBT S1. Fig. 15b shows the diagnosis result of open-circuit fault in S1 and S3, the order of which is (S3 fault) → (normal state) → (S1 fault) → (S1 and S3 faults) → (S3 fault), and the order of the diagnosis results is consistent with the previous analysis, finally, the open-circuit fault locations are in IGBTs S1 and S3. With the help of a mature fault diagnosis classifier, the fault diagnosis system can accurately complete faults protection and location within 20 ms, and the dependence of the proposed method on the fault model is greatly reduced.

The RFs fault diagnosis classifier with the transient synthetic features can accurately identify the fault location of a single IGBT. Also, when open-circuit faults occur in two or more IGBTs, the fault locations can also be accurately and comprehensively diagnosed by statistical analysis of the diagnostic results.

# 5 Conclusion

This study has found that the three-phase PWM rectifiers can usually maintain operation when open-circuit faults occur in IGBTs. By analysing the features of open-circuit faults in IGBTs of three-phase PWM rectifiers, it is found that the lower half period of the phase current will be affected when the open-circuit faults occur in upper IGBTs, and the upper half period will be affected when open-circuit faults occur in lower IGBTs.

Compared with the ANN and SVM algorithms, the performance of RFs algorithm is better in training time and diagnosis accuracy. Meanwhile, the performance of the multiplicative features is better than the others, and the accuracy of RFs fault diagnosis classifier is improved by synthetic features. What is more, the proposed method, a data-driven method, relies less on fault models of power electronic converters.

Finally, the online fault diagnosis experiments for three-phase PWM rectifier were carried out, the experimental results have been adopted to indicate the effectiveness of the proposed method. The proposed method can accurately locate the open-circuit faults in IGBTs while ensuring the system safety.

# 6 Acknowledgment

This research was funded by the National Key R&D Program of China under grant number 2017YFB0903300.